\documentclass[12pt,aps]{article}
\usepackage{amsmath}
\usepackage{amstext}
\usepackage{amsfonts}
\usepackage{graphicx,epsfig}
\usepackage{indentfirst}
\begin{document}

\title{The Feasibility of Testing LHVTs in Charm Factory}

\author{Junli Li$^{1}$\footnote{jlli04@mails.gucas.ac.cn} and
Cong-Feng Qiao$^{2,1}$\footnote{qiaocf@gucas.ac.cn}\\
\small$^{1}$ Dept. of Physics, Graduate
University, the Chinese Academy of Sciences  \\
\small {YuQuan Road 19A, 100049, Beijing, China}\\
\small$^2$ CCAST(World Lab.), P.O. Box 8730, Beijing 100080,
China}
\date{}
\maketitle

\begin{abstract}
It is commonly believed that the LHVTs can be tested through
measuring the Bell's inequalities. This scheme, for the massive
particle system, was originally set up for the entangled
$K^0\bar{K^0}$ pair system from the $\phi$ factory. In this Letter
we show that the $J/\Psi\rightarrow K^0\bar{K^0}$ process is even
more realistic for this goal. We analyze the unique properties of
$J/\Psi$ in the detection of basic quantum effects, and find that
it is possible to use $J/\Psi$ decay as a test of LHVTs in the
future $\tau$-Charm factory. Our analyses and conclusions are
generally also true for other heavy Onium decays.
\end{abstract}

\vspace{6pt}
%\section{introduction}%%%%%%%%%%%%%%%%%%%%%%%%%%%%%%%%%%%%%%%%%%%
In 1935, Einstein Podolsky and Rosen (EPR) \cite{epr} demonstrated
that quantum mechanics (QM) could not provide a complete
description of the ``physical reality'' for two spatially
separated but quantum mechanically correlated particle system.
Alternatively, local hidden variable theories (LHVTs) have been
developed to restore the completeness of QM. In 1964, Bell
\cite{bell} showed that in realistic LHVTs' two particle
correlation functions satisfy a set of Bell inequalities (BI),
whereas the corresponding QM predictions can violate such
inequalities in some region of parameter space. This leads to the
possibility for experimental test of the validity of LHVTs in
comparison with QM.

Many experiments in regard of the Bell inequalities have been
carried out yet by using the entangled photons, e.g. in Ref.
\cite{aspect}. And, the relevant progresses in this direction were
realized by using the Parametric Down Conversion (PDC) \cite{pdc}
technique to generate entangled photon pairs. As far, all these
experimental results \cite{aspect,pdc,weihs} are substantially
consistent with the prediction of standard QM.

Making a complete test of LHVTs, to go beyond the massless quanta,
the photons, is necessary. To this aim, the spin singlet state, as
first advocated by Bohm and Aharonov \cite{ab} to clarify the
argument of EPR, was also realized in experiment. Lamehi-Rachti and
Mitting \cite{wmitting} carried out the experiment in the low energy
proton-proton scattering using Saclay tandem accelerator. And, their
measurement of the spin correlation of protons gave as well a good
agreement with QM.

People notice that the previous experiments are mainly restricted
in the electromagnetic interaction regime, i.e., employing the
entangled photons, no matter they are generated from atomic
cascade or PDC. Considering of the fundamental importance of the
question, to test the LHVTs in other experiments with massive
quanta and different interaction nature is necessary \cite{abel}.
The meson pairs with strong or weak interaction in the high energy
region can possibly play a role in this aim.

As early as 1960s the EPR-like features of the $K^0\bar{K^0}$ pair
in decays of the $J^{PC}=1^{--}$ vector particles was noticed by
Lipkin \cite{lipkin}. The early attempts of testing LHVTs through
the Bell inequality in high energy physics focused on exploiting
the nature of particle spin correlations
\cite{abel,tqvist,privitera}. In Ref. \cite{tqvist} T\"ornqvist
suggested to measure the BI through
\begin{eqnarray}
e^+e^-\rightarrow\Lambda\bar{\Lambda}\rightarrow\pi^-\; p\;
\pi^+\; \bar{p}
\end{eqnarray}
process. A similar process was suggested by Privitera
\cite{privitera}, i.e.,
\begin{eqnarray}
e^+e^- \rightarrow \tau^+ \tau^- \rightarrow \pi^+
\bar{\nu}_{\tau} \pi^-\nu_{\tau}\; .
\end{eqnarray}
However, later on people found that such proposals have
controversial assumptions \cite{book}. Now we realize that in
testing the LHVTs in high energy physics, using the ``quasi-spin"
to mimic the photon polarization in the construction of entangled
states is a practical way.

The $B^0\bar{B^0}$ system which produced at the $\Upsilon(4S)$
resonance has been measured \cite{b0}. However, debate on whether
it was a genuine test of LHVTs or not is still on going
\cite{verified}. The $K^0\bar{K^0}$ system is an ideal tool to
investigate the LHVT problem. In this case, there has no
difficulties which B-meson system faces \cite{prl88}. It was
normally considered to test the LHVTs in $K^0\bar{K^0}$ system at
the $\phi$-factory. However, in the following we argue that, in
fact, the $K^0\bar{K^0}$ system generated from the heavy
quarkonium decay is even more realistic and suitable for this aim.

For Onium decaying into the $K^0\bar{K^0}$ pair, spin odd of the
initial particle is the only requirement for the K-meson pair to
form the entangled state under the assumption of CPT conservation
\cite{cpt}. Thus, any vector-like particle or resonance can be
used to construct the entangled state of $K^0\bar{K^0}$. To test
the LHVTs in kaon system through heavy quarkonium decay has two
distinct merits. First, the initial state is almost a pure state;
secondly, the heavy quarknoium mass enables the entangled states
with large spacial separation, which is easier for experimental
test as shown below.

Neglecting the small CP-violation effects, the initial
$K^0\bar{K^0}$ pair decay from $J/\psi$ can be written as:
\begin{eqnarray}
|J/\psi(0)\rangle=\frac{1}{\sqrt{2}}[K_SK_L-K_LK_S]\ ,\label{phi0}
\end{eqnarray}
where $K_S=(K^0+\bar{K^0})/\sqrt{2}$ and
$K_L=(K^0-\bar{K^0})/\sqrt{2}$ are the mass eigenstates of the $K$
mesons. The key point to use kaon system for testing the LHVTs is
to generate a nonmaximally entangled (asymmetric) state, as
proposed in Refs. \cite{prl89,quant}
\begin{eqnarray}
|J/\Psi(T)\rangle=\frac{1}{\sqrt{2+|R|^{\,2}}}\,
[K_SK_L-K_LK_S-re^{-i(m_L-m_S)\,T+\frac{\Gamma_S-\Gamma_L}{2}\,
T}K_LK_L]\ .\label{jpsiT}
\end{eqnarray}
Here, $r$ is the regeneration parameter to be the order of
magnitude $10^{-3}$ \cite{prl89}; $\Gamma_L$ and $\Gamma_S$ are
the $K_L$ and $K_S$ decay widths, respectively; T is the evolution
time of kaons after their production. Technically, this asymmetric
state can be achieved by placing a thin regenerator close to the
$J/\psi$ decay point \cite{prl88}.

Four specific transition probabilities for joint measurements from
QM take the following form:
\begin{eqnarray}
P_{QM}(K^0,\bar{K^0}) & \equiv & |\langle
K^0\bar{K^0}|J/\Psi(T) \rangle|^{\ 2}\nonumber \\
 & = & \frac{|\,2-re^{-i(m_L-m_S)\,T+\frac{\Gamma_S-\Gamma_L}
 {2}\,T}|^{\,2}}{4(2+|\,r|^{\,2}
 e^{(\Gamma_S-\Gamma_L)\,T})}\nonumber \\
 & = & \frac{|2+\mathrm{R}e^{i\varphi}|^{\,2}}
 {4(2+|R|^2)}\ ,
 \label{tp1}
\end{eqnarray}
\begin{eqnarray}
P_{QM}(K^0,K_L) & \equiv & |\langle K^0 K_L|J/\Psi
(T)\rangle|^{\ 2}\nonumber \\
& = & \frac{|\,1-re^{-i(m_L-m_S)\,T +
\frac{\Gamma_S-\Gamma_L}{2}\,T}|^{\,2}}{2(2 +
|\,r|^{\,2}e^{(\Gamma_S-\Gamma_L)\,T})}\nonumber \\
& = & \frac{|1 + \mathrm{R}e^{i\varphi}|^{\,2}}{2(2 + |R|^2)}\ ,
 \label{tp2}
\end{eqnarray}
\begin{eqnarray}
P_{QM}(K_L,\bar{K^0}) & \equiv & |
\langle K_L\bar{K_0}|J/\Psi(T)\rangle|^{\ 2}\nonumber \\
& = & \frac{|\,1 - re^{-i(m_L-m_S) \,T + \frac{\Gamma_S -
\Gamma_L}{2} \,T}|^{\,2}}
{2(2+|\,r|^{\,2}e^{(\Gamma_S-\Gamma_L)\,T})}\nonumber \\
& = & \frac{|1+\mathrm{R}e^{i\varphi}|^{\,2}}{2(2+|R|^2)}\ ,
 \label{tp3}
\end{eqnarray}
\begin{eqnarray}
P_{QM}(K_SK_S)& \equiv & |\langle K_SK_S|J/\Psi(T)\rangle|^{\
2}\nonumber \\ & = &\ 0 \ ,
 \label{tp4}
\end{eqnarray}
where $\mathrm{R}=-|R|=-|r|e^{\frac{\Gamma_S-\Gamma_L}{2}\,T}$,
$\varphi$ is the phase of $R$.
\begin{figure}[t,m,u]
\begin{center}
\includegraphics[width=9cm,height=8cm]{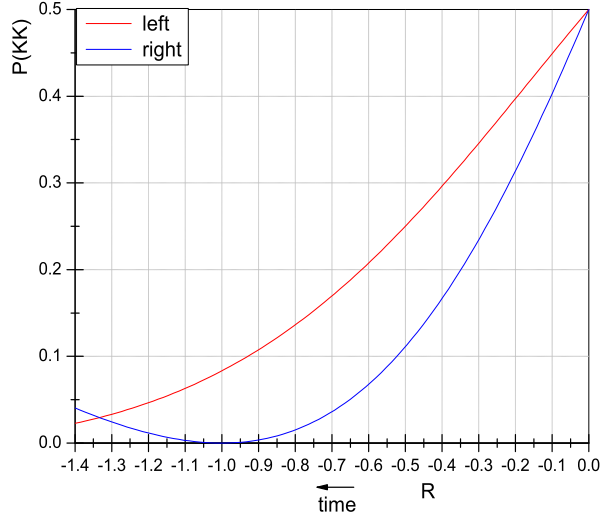}
\end{center}
\caption{\label{kk} \small The comparison of QM prediction with
the EI constraint. The horizontal axis refers to the values of R;
vertical axis refers to the probabilities of corresponding events.
The red line represents for the left side of inequality
(\ref{ei}), and the blue for the sums of the right three terms.}
\end{figure}

\begin{figure}
\begin{center}
\includegraphics[width=9cm,height=8cm]{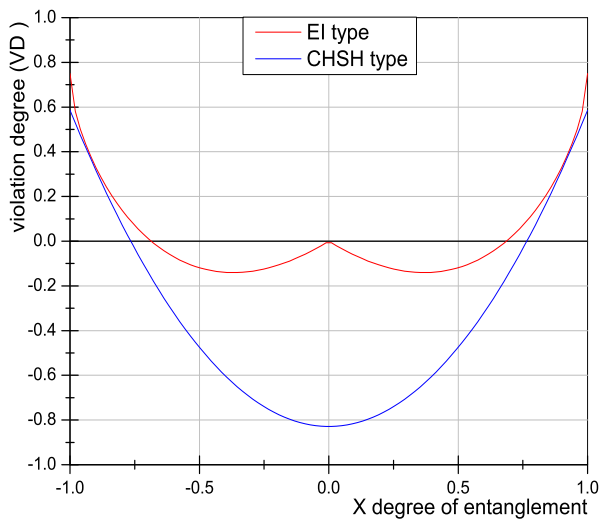}
\end{center}
\caption{\label{breaking1}\small The violation degree of the Bell
inequalities (red for EI type and blue for CHSH type) in terms of
the entanglement. Here, for the sake of transparency, we make a
coordinator exchange, the $C = 1-x^2$. The magnitudes of VD less
than zero means the broken of the BIs. }
\end{figure}

Eberhard's version of BI (EI) for kaon system takes the form
\cite{Eberhard,garuccio}:
\begin{eqnarray}
P_{LR}(K^0\bar{K^0}) & \leq &
P_{LR}(K^0,K_L)+P_{LR}(K_S,K_S)+P_{LR}(K_L,\bar{K^0})\ ,
\label{ei}
\end{eqnarray}
where $P_{\rm {LR}}$ denotes the LHVT transition probability with
the subscripts L and R symbolling the left and right kaons, and we
assume detection efficiency of the kaon decay products is one
hundred per cent. For the case of QM, substituting equations
(\ref{tp1}) - (\ref{tp4}) into the inequality (\ref{ei}) and
assuming $\varphi = 0$, we have
\begin{eqnarray}
\frac{(2+\mathrm{R})^2}{4(2+\mathrm{R}^2)} & \leq &
\frac{(1+\mathrm{R})^2}{2(2+\mathrm{R}^2)}\; +\; 0\; + \;
\frac{(1+\mathrm{R})^2}{2(2+\mathrm{R}^2)}\ . \label{ei1}
\end{eqnarray}
The above inequality is apparently violated by QM while
$\mathrm{R} = -1$. The R dependence of the violation of inequality
(\ref{ei1}) is plotted in the Figure 1. Taking concurrence $C$ as
a measure of entanglement \cite{concurrence} we have:
\begin{eqnarray}
C(J/\Psi)=|\langle
J/\Psi|\widetilde{J/\Psi}\rangle|=\frac{2}{2+|R|^{\,2}}=
\frac{2}{2+|r|^{\,2}e^{(\Gamma_S-\Gamma_L)\,T}}\ , \label{ev1}
\end{eqnarray}
where $|\widetilde{J/\Psi}\rangle = \sigma^1_y \sigma^2_y
|(J/\Psi)^{*}\rangle\ $ and $\sigma^{1,\ 2}$ are Pauli matrices.
$C$ changes between null to unit for no entanglement and full
entanglement. Superficially, equation (\ref{ev1}) shows that the
state become less entangled with the time evolution. However, our
result, as shown in Figure \ref{kk}, indicates clearly that there
exists a period of time during which the violation of the
inequality become larger with the time evolution. The larger
violation for non-maximally entangled state than that of the
maximally entangled state was also obtained by Ac\'{i}n {\it et
al.} \cite{twothree}, where they use two d-dimensional ($d\geq3$)
quantum systems. To clarify this phenomenon we express the
violation degree (VD) of the inequalities (left side subtract the
right side) in term of $C$ and compare it with the usual CHSH
inequality \cite{CHSH}. In Figure \ref{breaking1}, the different
VD behaviors of CHSH's and Eberhard's Inequalities are presented.
For CHSH case, the $VD_{\rm CHSH}$ is obtained in the same
condition as what the maximal violation happens in the full
entanglement, the $C = 1$. We have
\begin{eqnarray}
VD_{CHSH} = \sqrt{2}(1+C)-2\ .
\end{eqnarray}
In fact, the above $VD_{\rm CHSH}$ can be deduced from the results
given in Refs. \cite{gisin,kar,degree of two}. For EI case,
substituting Eq. (\ref{ev1}) into (\ref{ei1}) we have
\begin{eqnarray}
VD_{EI}=\frac{-3(1-C)+2\sqrt{2}\sqrt{C-C^2}}{4}\ .
\end{eqnarray}
Here,  in EI the counterintuitive quantum effect shows up, i.e.
the less entanglement corresponding to a larger VD in some region,
which is induced by the rotational symmetry breaking of the
initial singlet state in quasi-spin space.
\begin{figure}
\begin{center}
\includegraphics[width=10cm,height=8cm]{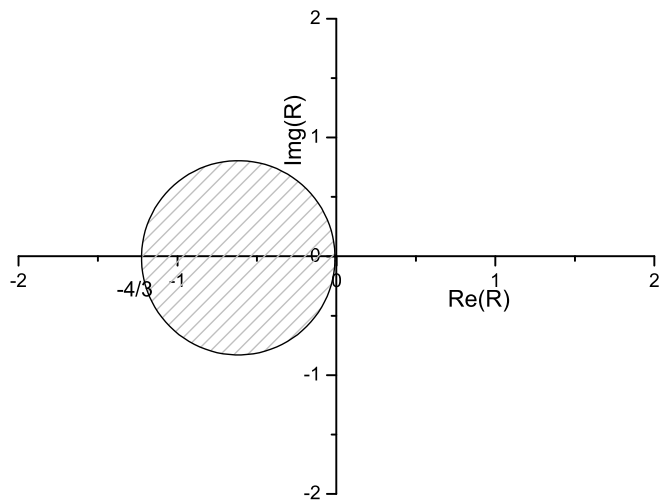}
\end{center}
\caption{\label{phase}\small The practical complex $R$ values in
distinguishing QM from LHVTs according to the inequality
(\ref{ei}).}
\end{figure}

With the time evolution, when $R$ becomes less than
$-\frac{4}{3}$, the QM and LHVTs both satisfy the inequality
(\ref{ei}). Given a certain asymmetrically entangled state, the
Hardy state \cite{hardy}, in the region of $R\in[-4/3,0)$ the QM
and LHVTs can be distinguished from the EI. In preceding
discussion, for simplicity we chose the phase of the complex
quantity $R$ to be zero. However, in practice, the phase $\varphi$
can have a non-zero magnitude. Generally, as shown in Figure
\ref{phase}, the EI is violated by the QM in the shaded region of
$R$.

To detect the $VD_{\rm EI}$, in experiment one needs to detect the
decay events taking place between $T$ and $T+\Delta T$. Here, $T$
is the time when the entangled states are asymmetric and kept to
be space-like in the follow up measuring interval $\Delta T$. In
order to make sure the misidentification of $K_S$ and $K_L$ is
less than one thousandth, the $\Delta \mathrm{T}$ should be larger
than $5.5\tau_S$, which can be easily obtained, like
\begin{eqnarray}
P_{survived}(K_L(\Delta T)) & = & \;|\,e^{-\frac{\Gamma_L}{2}
\Delta T-im_L\Delta T}|^{\ 2}\; \simeq 99.0\%\ ,
\end{eqnarray}
\begin{eqnarray}
P_{decayed}(K_S(\Delta T)) & = & 1- |\,e^{-\frac{\Gamma_S}{2}
\Delta T-im_S\Delta T}|^{\ 2}\simeq 99.6\%\ .
\end{eqnarray}

In comparison with $\phi$ factory, the kaon system produced from
the heavy Onium decay is easier to be space-like. The kaon pairs
generated from $\phi$ decays moving at $\beta_K  \simeq 0.22$. To
ensure the entangled kaons to be space-like, the evolution time
$T$ should larger than $11\tau_S$, which narrows the available
region of $R$ in discriminating QM from LHVTs. In this case, the
upper bound for $R$ is roughly
\begin{eqnarray}
|R|=|r|e^{\frac{\Gamma_S}{2} T} \sim 0.4\ .
\end{eqnarray}
For kaons from $J/\Psi$ decay, $\beta_K=\frac{v_K}{c} \simeq
0.94$. This implies that as long as
$T>(\beta^{-1}-1)\Delta\mathrm{T}/2\simeq0.15\tau_S$ the
space-like separation is guaranteed. Correspondingly, the $R$ is
pretty small:
\begin{eqnarray}
|R|=|r| e^{\frac{\Gamma_S}{2} T} \sim 10^{-3}\ .
\end{eqnarray}
This unique property of kaon system from Charmonium decays enables
us to observe a peculiar QM effect, that is the less entangled
state lead to a larger BI violation.

%%%%%%%%%%%%%%%%%%%%%%%%%%%%%%%%%%%%%%%%%%%%%%%%%%%%%%%%%%%%%%%%%
In conclusion, in this paper we propose to use the entangled kaon
system produced in heavy quarkonium decay to test the LHVTs, that
is the violation of a certain kind of Bell Inequality. We compare
the difference in $\phi$ factory and quarkonium decays, and find
that it is even possible to distinguish the Quantum Mechanics from
the Local Hidden Variable Theories in the charmonium decays. In
the charmonium decays the entangled kaon state is more energetic
and easier to be space-like. On the other hand, in charmonium
decays it is possible to observe a peculiar Quantum effect, that
is the decreasing entanglement corresponds to the increasing BI
violation in certain kinematic region. To be noticed that in order
to have a unitary time evolution for the kaon system one has to
include their decay states. This feature affects the Bell
inequalities in certain degree, in particular the CHSH inequality
\cite{bertlmann}, however, is not taken into account. In this
work, we also give out the complete $R$ values for the BI
violation in kaon system and express the violation in terms of
degree of entanglement for the first time. Although the
experimental efficiency for detecting the kaon system in heavy
Onium decays is lower than the efficiency in $\phi$ factory,
whereas our calculation shows that it is still possible to use
$J/\Psi$ decay to kaons as a test of LHVTs in the future high
statistic $\tau$-Charm factory. Moreover, it should be noted that
our analyses and conclusions are generally also true for other
heavy vector-like Onium decays.
%%%%%%%%%%%%%%%%%%%%%%%%%%%%%%%%%%%%%%%%%%%%%%%%%%%%%%%%%%%%%%%%

This work was supported in part by the Nature Science Foundation
of China and by the Scientific Research Fund of GUCAS(NO.
055101BM03). We are grateful to Z.P.Zheng, X.Y.Shan and H.B.Li for
their helpful discussion.

\end{document}